\documentclass[cameraready]{Interspeech}
\usepackage[table]{xcolor}   

\usepackage{amsmath,graphicx, multirow, makecell}
\usepackage{float, booktabs, xcolor}
\usepackage{amssymb}
\usepackage{pifont}
\usepackage{soul} 
\usepackage{stmaryrd}
\usepackage{trimclip}

\makeatletter
\DeclareRobustCommand{\shortto}{%
  \mathrel{\mathpalette\short@to\relax}%
}

\newcommand{\short@to}[2]{%
  \mkern2mu
  \clipbox{{.2\width} 0 0 0}{$\m@th#1\vphantom{+}{\shortrightarrow}$}%
  }
\makeatother

\def\L{{\cal L}}

\usepackage{graphicx}  
\definecolor{lightgray}{gray}{0.9}
\title{SEMamba++: A General Speech Restoration Framework Leveraging Global, Local, and Periodic Spectral Patterns}

\author[affiliation={1}]{Yongjoon}{Lee}
\author[affiliation={1}, correspondingauthor]{Jung-Woo}{Choi}

\usepackage{comment}

\address{
    $^1$ Korea Advanced Institute of Science and Technology (KAIST), Daejeon, Korea 
}

\email{yongjoonlee@kaist.ac.kr, jwoo@kaist.ac.kr}


\begin{document}

\maketitle

\begin{abstract}
General speech restoration demands techniques that can interpret complex speech structures under various distortions. While State-Space Models like SEMamba have advanced the state-of-the-art in speech denoising, they are not inherently optimized for critical speech characteristics, such as spectral periodicity or multi-resolution frequency analysis. In this work, we introduce an architecture tailored to incorporate speech-specific features as inductive biases. In particular, we propose the Global, Local, and Periodic (GLP) module, a frequency feature extraction block that effectively and efficiently leverages the properties of frequency bins. Then, we design a multi-resolution parallel time-frequency dual-processing block to capture diverse spectral patterns, and a learnable mapping to further enhance model performance. With all our ideas combined, the proposed SEMamba++ achieves the best performance among multiple baseline models while remaining computationally efficient.

\end{abstract}
 
%
\section{Introduction}
\label{sec:intro}
General speech restoration (GSR)~\cite{Liu_2022} refers to the task of recovering high-quality speech from signals affected by a range of degradations, such as noise, reverberation, bandwidth limitation, and clipping. 
This problem is particularly relevant in real-world environments, where speech is often captured under adverse acoustic conditions or with limited recording equipment, resulting in multiple and overlapping types of distortion. 
Unlike speech denoising or dereverberation that aims to \textit{predict} clean speech by removing noise or reverberation, GSR not only predicts clean speech but also \textit{generates} the missing speech fragment so that the resulting speech is perceptually natural. This fragment may be high-frequency bands in bandwidth-limited situations or a high-amplitude signal in clipped scenarios. 
This hybrid nature has led to various approaches to solving GSR. 

Generative methods have been used for their high perceptual quality and versatility in generating missing speech fragments. Universe~\cite{serrà2022universalspeechenhancementscorebased} introduced score-based diffusion~\cite{song2021scorebasedgenerativemodelingstochastic} to solve GSR, on which Universe++~\cite{scheibler2024universalscorebasedspeechenhancement} improved using a hybrid of score-matching~\cite{song2021scorebasedgenerativemodelingstochastic} and generative adversarial networks (GAN)~\cite{goodfellow2014generativeadversarialnetworks}. ANYENHANCE ~\cite{Zhang_2025} is a multi-task masked generative model (MGM)~\cite{lezama2022improvedmaskedimagegeneration} utilizing prompt-guidance and self-critic mechanisms to successfully solve speech enhancement. In generative methods, language models (LMs) have also emerged as a viable option due to their generalization capacity. MaskSR~\cite{li2024masksrmaskedlanguagemodel} is a masked language model (MLM) capable of restoring full-band speech with discrete acoustic tokens from pretrained neural audio codecs. LLaSE-G1 ~\cite{kang2025llaseg1incentivizinggeneralizationcapability} addresses multi-task GSR utilizing a language model. However, LM-based methods require substantial amounts of training data. 

Surprisingly, discriminative methods, although thought to be less perceptually natural and versatile than generative methods, yield impressive results in GSR. Discriminative methods~\cite{scalingbeyonddenoising, chao2025universalspeechenhancementregression} achieved high performance in GSR challenges such as URGENT~\cite{saijo2025interspeech2025urgentspeech} and AATC~\cite{zhang2025ccfaatc2025speech}. This success can be attributed to architectural improvements and advanced training techniques (e.g., improved loss functions) inherited from speech denoising. CMGAN~\cite{Cao_2022} introduced a Conformer~\cite{gulati2020conformerconvolutionaugmentedtransformerspeech}-based architecture with dilated DenseNets~\cite{9054536}. CMGAN used a metric discriminator~\cite{fu2019metricgangenerativeadversarialnetworks} with reconstruction losses to directly optimize perceptual metrics. MP-SENet~\cite{Lu_2023} extended this framework by introducing parallel magnitude and phase decoding. More recently, SEMamba~\cite{chao2025investigationincorporatingmambaspeech} replaced the Conformer with Mamba~\cite{gu2024mambalineartimesequencemodeling}, leveraging its selective state-space mechanism. 
For the aforementioned methods, separate feature extraction for time and frequency bins has been a key design; a procedure referred to as time-frequency dual-path (TFDP) processing in this work. Unlike the image domain, where the height and width represent the same characteristics at different locations, the time and frequency bins in a speech spectrum exhibit heterogeneous properties. To this, feature extraction across time and frequency bins has been implemented separately.

Although time and frequency features are processed separately, both are often processed with modules of the same architecture~\cite{Cao_2022,chao2025investigationincorporatingmambaspeech, kühne2025xlstmsenetxlstmsinglechannelspeech}. 
To better reflect the distinction between time and frequency features, the design should be tailored to address a specific domain. In particular, frequency feature extraction modules need to capture global, local, and periodic patterns of the speech spectrum effectively. 
Conformer~\cite{Cao_2022} and SpatialNet~\cite{10423815}-style frequency feature extraction have been popular choices for extracting global and local features by jointly using full-band (global) modules and sub-band (local) modules. 
However, this frequency feature mixing may not be optimal for GSR due to the lack of local-global selectivity and the spectral periodicity modeling capacity.
First, the serially connected local and global modules reduce selectivity: the model’s ability to prioritize either local or global frequency representations depending on the degradation characteristics. This selectivity between local and global representations is critical, as degraded speech in GSR exhibits distinct characteristics across different degradation kernels. For example, the global branch should be prioritized for bandwidth extension, where the distortion is caused by the truncation of a wide spectral region. 
Moreover, the periodicity in frequency has also been underexplored, despite the importance of near-harmonic structures from vocal excitations. In this regard, a means to focus on global, local, and periodic feature modeling is required. 

\begin{figure}[htbp]
    \centering
    \includegraphics[width=0.8\columnwidth]{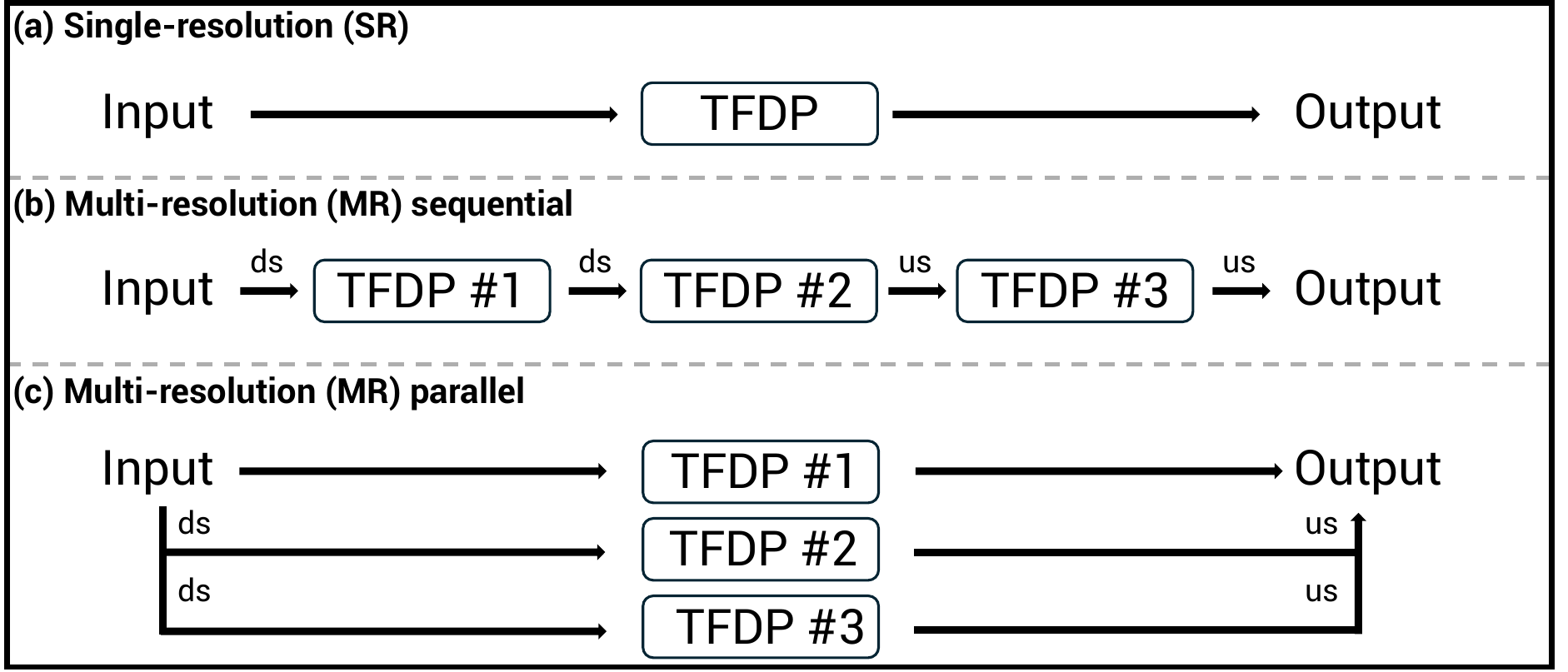}
    \caption{Various TFDP processing methods. ds and us denote downsampling and upsampling, respectively.}
    \label{fig:abc}
\end{figure}

Aside from the frequency feature extraction methods, the single-resolution TFDP processing may not be optimal in GSR. TFDP is often implemented at a single resolution—without aggressive down- or upsampling—to preserve fine details and minimize upsampling artifacts~\cite{pons2021upsamplingartifactsneuralaudio}. Since this design entails a high number of TF bins, the models are necessarily equipped with modules capable of modeling long sequences (or many bins)~\cite{Cao_2022, chao2025investigationincorporatingmambaspeech,lee_2024,kühne2025xlstmsenetxlstmsinglechannelspeech}. 
Single-resolution processing faces two limitations. 
First, scaling model capacity via deeper or wider architectures incurs excessive computational overhead due to the long sequence length. A recent work~\cite{scalingbeyonddenoising} resolves this problem by utilizing progressive block expansion strategies. This approach enables efficient training of high-capacity TFDP models, but the training strategy is complex and does not alleviate the heavy computations during inference, which is critical in resource-restrained settings. 
Second, single-resolution processing misses opportunities for multi-scale feature extraction. For instance, in bandwidth-limited signals, information from low-frequency bins can be better propagated to high-frequency bins when downsampling is applied along the frequency axis. A recent study~\cite{kuhne2026exploringresolutionwisesharedattention} uses multi-resolution sequential TFDP processing to resolve this issue, though the sequential TFDP processing does not fully exploit the diverse speech spectral patterns since TFDP-processed output from \#1 affects TFDP processing at \#2 and \#3 in Figure~\ref{fig:abc}. 


To resolve the aforementioned issues, we first propose Frequency GLP (Global, Local, and Periodic), a novel frequency-processing block that employs a parallel connection of the global periodicity (GP) module and the local (L) module. The GP module comprises a series of Fourier Analysis Network (FAN)~\cite{dong2025fanfourieranalysisnetworks} applied directly to the frequency bins, whereas the L module comprises a series of convolutional blocks. The method effectively leverages the intrinsic properties of the speech spectrum, achieving the best performance over a range of advanced baselines. Note that FAN has been applied in GSR~\cite{scalingbeyonddenoising} to replace the Multi-Layer Perceptron (MLP), but the direct application on frequency bins to capture the periodicity in spectral patterns has been underexplored. 

Second, motivated by multi-resolution parallel processing in the image domain~\cite{zamir2020learningenrichedfeaturesreal}, we design a multi-resolution parallel TFDP processing with frequency-only downsampling. Our architecture operates on three frequency resolutions while preserving the temporal resolution. By downsampling only along the frequency axis, the architecture enables efficient performance elevation without sacrificing temporal fidelity. To further capture diverse spectral patterns in degraded speech, we conduct parallel TFDP processing for each resolution. As illustrated in Figure~\ref{fig:abc}, this strategy allows each resolution to specialize in distinct spectral patterns, since TFDP-processed results from \#1 do not affect TFDP processing at \#2 or \#3. Related to our work, DCCRN~\cite{hu2020dccrndeepcomplexconvolution} explored the U-Net architecture with frequency-only downsampling, but its feature processing is primarily focused on the most compressed resolution, and empirical comparisons across different downsampling strategies have been underexplored. Overall, our contribution can be summarized as follows: 

\begin{itemize}

\item We propose Frequency GLP, a novel frequency processing module that effectively and efficiently captures global, local, and periodic frequency patterns. This improves restoration quality in both in-domain and out-of-domain scenarios. 

\item We design multi-resolution parallel TFDP processing with frequency-only downsampling. We demonstrate that parallel processing allows the model to capture diverse patterns in the speech spectrum.

\item We propose a learnable softplus-based mapping that assigns frequency-wise hyperparameters to leverage the distinct properties of frequency bins.

\end{itemize}

\section{Preliminary}
\subsection{Fourier Analysis Network}

Recently, FAN~\cite{dong2025fanfourieranalysisnetworks} has emerged as a promising alternative to the MLP in modern deep learning models. FAN utilizes Fourier analysis to promote the periodicity in data. Given the input $x \in \mathbb{R}^{d_{x}}$, FAN is defined as follows:
\begin{equation}
\phi(x) \triangleq 
\bigl[\, \cos(xW_p) \Vert \sin(xW_p) \Vert \sigma(B_{\bar{p}} + xW_{\bar{p}}) \,\bigr],
\end{equation}
where $\phi(x)$, $\sigma(\cdot)$, and $\big[\cdot \Vert \cdot\big]$ refer to the FAN operation, standard activation (e.g., Gaussian Error Linear Unit (GELU)), and concatenation, respectively. $W_p \in \mathbb{R}^{d_x \times d_p}$, $W_{\bar{p}} \in \mathbb{R}^{d_x \times d_{\bar{p}}}$, $B_{\bar{p}} \in \mathbb{R}^{d_{\bar{p}}}$, and FAN operation outputs $\phi(x)\in \mathbb{R}^{2d_p+d_{\bar{p}}}$. 
The FAN layer balances periodicity and general-feature modeling through the feature sizes $d_p$ and $d_{\bar{p}}$. In addition, it is necessary to apply at least one linear layer after FAN so that the Fourier coefficients can be learned explicitly. 
Since its sine and cosine activations share the same parameters $W_p$, FAN can have a smaller parameter size than that of a vanilla linear layer with the same output dimension. 
\begin{figure*}[t]
    \centering
    \includegraphics[width=1.\textwidth]{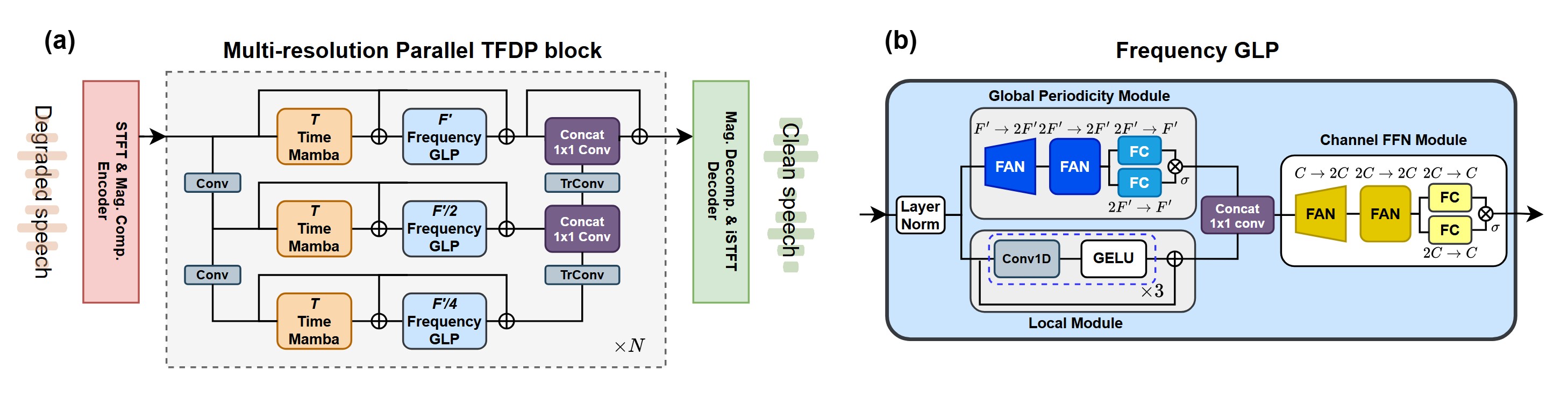}
    \caption{Overall architecture of SEMamba++ (a) and Frequency GLP (b). In (a), "Mag.", "Comp.", and "Decomp." refer to magnitude, compression, and decompression, respectively. "Conv" and "TrConv" indicate the 2d convolution and transposed convolution that down- and upsample along the frequency axis. (b) describes Frequency GLP with frequency dimension $F'$.}
    \label{fig:modelarch}
    \vspace{-0.5em}
\end{figure*}

\section{Methodology}
\label{sec:method}

The proposed model, illustrated in Figure~\ref{fig:modelarch}(a), has an encoder-bottleneck-decoder architecture. Its encoder and decoder design is inherited from~\cite{Lu_2023}. In short, the degraded speech is converted into the magnitude and phase spectrum using the Short-Time Fourier Transform (STFT), followed by power-law compression~\cite{Lu_2023} of the magnitude spectrum. The spectra are then fed to the dilated DenseNet~\cite{9054536}-based encoder, which expands the channel dimension to $C$ and halves the number of frequency bins to $F'=F/2$, where $C$ and $F$ denote the number of channels and frequency bins, respectively. We have $N$ bottleneck blocks to perform TFDP processing, where we set $N$ to 4 and $C$ to 48. After the bottleneck blocks, the features are passed through two dilated DenseNet-based decoders to estimate the magnitude and phase components of clean speech in parallel. Particularly, the number of frequency bins is doubled to $F=2F'$ in the decoder. The magnitude is power-law-decompressed, followed by the inverse STFT to generate a clean waveform. All source codes and demos are available\footnote{\url{https://sites.google.com/view/semambapp}}.

\subsection{Frequency GLP}
\label{sec:freqmix}

Frequency GLP, shown in Figure~\ref{fig:modelarch}(b), is the core module introduced for frequency feature extraction. We explain Frequency GLP by describing its two key aspects. 
First, the model comprises a parallel connection of the GP module and the L module, each addressing the full frequency band and subbands. After each module aggregates spectral features, 
the concatenation followed by a pointwise convolution serves as a selection operator to adjust the information flow from both modules and restore a speech fragment with a given degradation. The features are fed into a channel feedforward network (FFN) module to enhance the expressivity of frequency processing. 

The second main component of Frequency GLP is the FAN-based GP module. In the GP module, FAN is applied directly to the frequency axis rather than to the channel axis. Specifically, the linear layer is directly applied to frequency bins of dimension $F_{\mathrm{eff}}$, after which the sine and cosine activations follow. $F_{\mathrm{eff}}$ refers to the effective number of frequency bins in each resolution illustrated in Figure~\ref{fig:modelarch}(a). This enables the module to learn periodicity in the frequency bins, such as harmonics, via a Fourier series approximation. 
The GP module first employs a FAN layer to expand the feature dimension from $F_{\mathrm{eff}}$ to $2F_{\mathrm{eff}}$, followed by another FAN layer that maintains the dimension at $2F_{\mathrm{eff}}$. Finally, a linear layer-based gating is applied to learn the Fourier coefficients. The L module, consisting of a series of 1D convolutions with a kernel size of 3 and GELU activation, operates along the frequency axis to capture local relationships within subbands. 
The local module is beneficial in capturing spectral relationships that cannot be captured by the FAN-based GP module.
Next, the architecture of the channel FFN module is the same as that used in the GP module, but with a transposed input so that the channel dimension $C$ serves as the input to the linear layer. This design homogeneity is not only for simplifying the model architecture but also due to the finding~\cite{dong2025fanfourieranalysisnetworks} that applying FAN to latent features also helps capture periodicity. We set $d_p$ to $C/4$ for both the GP module and the Channel FFN module throughout our experiments.


\subsection{Multi-resolution parallel TFDP block}
\label{sec:multibranchTF}

We design a multi-resolution parallel TFDP block. Unlike single-resolution TFDP or multi-resolution sequential TFDP in Figure~\ref{fig:abc}, we stack multiple TFDP modules within a single block in parallel to process features of multiple resolutions, as illustrated in Figure~\ref{fig:modelarch}(a). We downsample frequency bins by a factor of $2^{r}$ where $r=1,2,..., R-1$ for each resolution, and $R$ denotes the number of resolutions in a multi-resolution parallel TFDP block. Parallel processing allows each branch to process the same signal at a different scale independently, so they learn complementary features. The frequency downsampling and upsampling are done by strided convolutions and transposed convolutions with a kernel size of (3, 4) and a stride of (1, 2), respectively, where the dimensions are ordered as $(T, F)$, followed by instance normalization~\cite{ulyanov2017instancenormalizationmissingingredient} and parametric ReLU (Rectified Linear Unit)~\cite{he2015delvingdeeprectifierssurpassing}. $T$ denotes the number of time bins. Note that the downsampling and upsampling operations retain the same hidden dimension $C$. TFDP processing comprises Time Mamba~\cite{chao2025investigationincorporatingmambaspeech} and Frequency GLP, with the expansion factor set to 2 in Time Mamba. After the TFDP processing, the bottom resolution (with frequency dimension equal to $F'/4$) is first merged with the middle resolution through channel-wise concatenation and a pointwise convolution. The resulting features are then fused with the top resolution outputs in the same manner. While drawing architectural inspiration from MIRNet~\cite{Zamir2022MIRNetv2}, our work departs from it by conducting an analytical study in Section~\ref{sec:mranalysis} on the complementary feature extraction capacity of the parallel TFDP processing, which was underexplored in the original framework. Furthermore, when combined with Frequency GLP, the proposed frequency-only downsampling exhibits significant efficiency, as it reduces the computational complexity of the FAN operation quadratically with respect to the effective dimension $F_{\mathrm{eff}}$.

\subsection{Learnable softplus mapping} 
Unlike conventional masking-based magnitude decoders, we employ a mapping-based magnitude decoder. Even though masking methods have demonstrated strong performance in denoising~\cite{chao2025investigationincorporatingmambaspeech, Lu_2023}, they show inconsistent performance in bandwidth extension, where bandlimited signals have zero energy in the upper frequency components. Inspired by the learnable masking in MetricGAN+~\cite{fu2021metricganimprovedversionmetricgan}, we design a learnable softplus as a mapping function. To capture frequency band-specific patterns, the model learns a distinct parameter $\beta_f$ for each frequency band $f \in \{1,2,\dots,F\}$.
The learnable softplus, for the input $x$ and the activation output $y$, is defined as
\begin{equation}
y_ = \frac{1}{\beta_f}\log\left(1 + e^{\beta_f x}\right),
\label{eq:learnable_softplus}
\end{equation}

\subsection{Vocoder-style training objective} 

While MetricGAN methods~\cite{chao2025investigationincorporatingmambaspeech, Lu_2023} effectively optimize the Perceptual Evaluation of Speech Quality (PESQ)~\cite{941023}, training with PESQ as the only adversarial objective may bias the model toward focusing only on enhancing PESQ instead of capturing the full aspects of perceptual quality~\cite{manocha2022audiosimilarityunreliableproxy}.
To address this, we adopt Least Squares GAN (LSGAN)~\cite{mao2017squaresgenerativeadversarialnetworks}, a widely validated adversarial loss in neural vocoding~\cite{lee2023bigvganuniversalneuralvocoder} and bandwidth extension~\cite{lee2025waveumambaendtoendframeworkhighquality}. LSGAN does not approximate an explicit objective quality score (e.g., PESQ) and lets the generator learn a more generalized notion of perceptual quality, which is empirically validated in Table~\ref{tab:thoroughablation}. Moreover, recent work demonstrates that LSGAN can encourage deterministic, clean waveform prediction from degraded inputs~\cite{babaev2024finallyfastuniversalspeech}. We use Multi-Scale Sub-Band Constant Q Transform Discriminator (MS-SB-CQTD)~\cite{gu2023multiscalesubbandconstantqtransform} and Multi-Resolution Discriminator (MRD)~\cite{jang2021univnetneuralvocodermultiresolution} as our discriminators. For MS-SB-CQTD, we used three sub-discriminators, setting the number of octaves to 8 and the number of bins per octave to 12, 24, and 36, respectively. The remaining settings for MS-SB-CQTD and MRD follow the approach in \cite{lee2023bigvganuniversalneuralvocoder}. The adversarial training objective is formulated as
\begin{align}
\mathcal{L}_{\mathrm{adv}}(G;D) &= \mathbb{E}_{x} \left[ (D(G(x)) - 1)^2 \right] \label{eq:lsgan_LG} \\
\mathcal{L}_{\mathrm{adv}}(D;G) &= \mathbb{E}_{y} \left[ (D(y) - 1)^2 \right] 
+ \mathbb{E}_{x} \left[ D(G(x))^2 \right], \label{eq:lsgan_LD}
\end{align}

\noindent where $y$ and $x$ represent the clean and degraded waveforms, respectively. $D$ and $G$ represent the discriminator and the generator, respectively. To stabilize the training, we employ several reconstruction losses. We use spectrogram magnitude L1 loss $\L_{\mathrm{mag}}$, anti-wrapping phase loss~\cite{ai2023neuralspeechphaseprediction} $\L_{\mathrm{awp}}$, consistency loss~\cite{zadorozhnyy2022scpganselfcorrectingdiscriminatoroptimization} $\L_{\mathrm{con}}$, complex loss~\cite{Cao_2022} $\L_{\mathrm{RI}}$, multi-scale mel spectrogram loss~\cite{kumar2023highfidelityaudiocompressionimproved} $\L_{\mathrm{mel}}$, and feature matching loss~\cite{larsen2016autoencodingpixelsusinglearned} $\L_{\mathrm{FM}}$. For $\lambda_{\mathrm{mag}}$, $\lambda_{\mathrm{awp}}$, $\lambda_{\mathrm{con}}$, and $\lambda_{\mathrm{RI}}$, we followed the hyperparameters introduced in the previous work~\cite{chao2025investigationincorporatingmambaspeech}. We set $\lambda_{\mathrm{mel}}=0.1$ and $\lambda_{\mathrm{FM}}=1.0$. Our discriminator training objective is $\mathcal{L}_D=\mathcal{L}_{\mathrm{adv}}(D;G)$, and the generator training objective is formulated as below:
\begin{align}
\mathcal{L}_{G}=\lambda_{\mathrm{adv}}\mathcal{L}_{\mathrm{adv}}(G;D) + \lambda_{\mathrm{mag}}\L_{\mathrm{mag}} + \lambda_{\mathrm{awp}}\L_{\mathrm{awp}} + \\ \lambda_{\mathrm{con}}\L_{\mathrm{con}} + \lambda_{\mathrm{RI}}\L_{\mathrm{RI}} + \lambda_{\mathrm{mel}}\L_{\mathrm{mel}} + \lambda_{\mathrm{FM}}\L_{\mathrm{FM}}
\end{align}

\section{Experimental settings}
\label{ssec:experiments}
 
\subsection{Training}
\label{sssec:training}

\textbf{Datasets} We utilized VCTK (version 0.92)~\cite{vctk} for speech data, except for speakers in p280 and p315 due to technical issues. For noise data, we used WHAM!~\cite{Wichern2019WHAM} and DNS Challenge 2020~\cite{reddy2020interspeech2020deepnoise}. WHAM! consists of various recordings from urban locations, whereas the DNS challenge 2020 noise dataset comprises sound clips from YouTube videos. To simulate reverberation, we used Arni~\cite{prawda2022calibrating} and the DNS5 challenge~\cite{7953152}. Data were sampled at 16 kHz and split into VCTK-GSR \textit{train} and VCTK-GSR \textit{test}, following~\cite{kuleshov2017audiosuperresolutionusing}. Note that the noise and room impulse response sources were also split into train-test subsets. 

\noindent\textbf{Degradation simulation} We considered noise, reverberation, bandwidth limitation, and clipping for distortion simulation. In training, we randomly applied various distortions. The signal-to-noise ratio (SNR) ranged from -10 to 20 dB, covering extremely noisy to clean conditions. Bandwidth limitation is applied at multiple cutoff frequencies (2–7 kHz) using various filter types (e.g., Chebyshev Type 1, Butterworth) and orders (2–8) to achieve diverse spectral effects.

\noindent\textbf{Training configurations} A segment size of 24,000 samples (1.5 seconds) was used for training. For STFT, we used a Hann window with a window length of 400 samples and a hop size of 100 samples. For the proposed method and variants in ablation studies, we set the learning rate as 2e-4 and the AdamW optimizer with $\beta_{1}$ and $\beta_{2}$ set to 0.8 and 0.99, respectively. The batch size was 8, and training was done for 100 epochs. An exponential learning rate decay with $\gamma=0.99$ was used. 

\begin{table*}[ht!]
\caption{Objective evaluation results on VCTK-GSR \textit{test}, URGENT 2025 \textit{val}, and URGENT 2025 \textit{test}. Reg. denotes the training with regressive loss. Official checkpoints have been utilized for models with $^*$.}
\label{tab:main-results}
\centering
\scriptsize
\setlength{\tabcolsep}{0.8pt}
\renewcommand{\arraystretch}{0.94}
\begin{tabular}{lcc
ccccc|ccccccccccc}
\toprule
& & 
\multicolumn{6}{c}{In-domain} 
& \multicolumn{9}{c}{Out-of-domain} \\
\cmidrule(lr){3-8} \cmidrule(lr){9-17}
\multirow{4}{*}{Model} & \multirow{4}{*}{Method} 
& \multicolumn{6}{c}{VCTK-GSR \textit{test}}
& \multicolumn{6}{c}{URGENT 2025 \textit{val}} & \multicolumn{3}{c}{URGENT 2025 \textit{test}}   \\
\cmidrule(lr){3-8} \cmidrule(lr){9-14} \cmidrule(lr){15-17}
&& \multicolumn{3}{c}{Perceptual quality} & \multicolumn{3}{c}{Signal fidelity} & \multicolumn{3}{c}{Perceptual quality} & \multicolumn{3}{c}{Signal fidelity} & \multicolumn{3}{c}{Perceptual quality} \\
\cmidrule(lr){3-5}\cmidrule(lr){6-8} \cmidrule(lr){9-11}\cmidrule(lr){12-14} \cmidrule(lr){15-17}
& & 
$\text{SCOREQ}^{\uparrow}$ & $\text{UTMOS}^{\uparrow}$ &  $\text{OVRL}^{\uparrow}$ & $\text{PESQ}^{\uparrow}$ & $\text{LSD}^{\downarrow}$ & $\text{LPS}^{\uparrow}$
& $\text{SCOREQ}^{\uparrow}$ & $\text{UTMOS}^{\uparrow}$ & $\text{OVRL}^{\uparrow}$ & $\text{PESQ}^{\uparrow}$ & $\text{LSD}^{\downarrow}$ & $\text{LPS}^{\uparrow}$ & $\text{SCOREQ}^{\uparrow}$ & $\text{UTMOS}^{\uparrow}$ & $\text{OVRL}^{\uparrow}$\\
\midrule
MP-SENet & GAN 
& 1.87 & 2.17
 & 2.82 & \underline{2.01} & 1.22      
& 0.42 & 1.59 & 1.90 & 2.88& \underline{1.63} & 1.78 & 0.59
& 1.38 & 1.71 & 2.81 \\

SEMamba & GAN 
& 1.89 & 2.34 
 & 2.88 & \textbf{2.13}& \textbf{1.19} & \textbf{0.45} 
& 1.66 & 2.06 & 2.92 & \textbf{1.67} & 1.83 & \underline{0.60} & 1.51 & 1.88 & 2.84
\\ 


USEMamba & Reg. 
& 1.87 & 2.30
& \underline{2.94} & 1.90 & \underline{1.21} & 0.42
&  1.77 & 2.05 & \underline{2.98} & 1.53 & \underline{1.51} 
& \underline{0.60} & 1.60 & 1.88 & \underline{2.91} \\

Universe++ & Score
& \underline{2.32} & \underline{2.74} 
& 2.74 & 1.45 & 1.34 & 0.22
& 1.26 & 1.78 & 2.45 & 1.18 & 2.13  & 0.17 & 1.20 & 1.69 & 2.38\\ 
Universe++$^*$ & Score 
& 1.77 & 2.06 
& 2.43 & 1.35 & 1.61 & 0.19
& 1.51 & 1.85 & 2.53 & 1.25 & 1.79 & 0.38
& 1.47 & 1.78 & 2.49 \\

LLaSE-G1$^*$ & LM
& 2.30 & 2.40 
& 2.47 & 1.20 & 2.62 & 0.26
& \underline{2.17} & \underline{2.11} & 2.69 & 1.15 & 1.93 & 0.54
& \underline{2.03} & \underline{1.94} & 2.62 \\
\midrule
SEMamba++ (Ours) & GAN 
& \textbf{3.27} & \textbf{3.55}
& \textbf{3.14} & 1.77 & 1.25 & \textbf{0.45}
& \textbf{2.67} & \textbf{2.82} & \textbf{3.20}  & 1.51 & \textbf{1.49} 
& \textbf{0.61} & \textbf{2.49} & \textbf{2.61} & \textbf{3.13} \\  

\bottomrule
\end{tabular}
\end{table*}

\subsection{Evaluation}
\label{sec:evaluation}
\textbf{Datasets} For in-domain performance evaluation, we used the VCTK-GSR \textit{test} dataset. Separately, to check generalization capacity to unseen data domains and degradations, we utilized the official validation set and blind test set in URGENT Challenge 2025~\cite{saijo2025interspeech2025urgentspeech}, the DNS 2020 test set~\cite{reddy2020interspeech2020deepnoise}, and the test set of AATC Challenge 2025~\cite{zhang2025ccfaatc2025speech}. The URGENT validation dataset (URGENT 2025 \textit{val}) employs a wider range of degradation types and languages.\footnote{Although approximately 10\% of the source recordings overlap, we regard the URGENT validation set as out-of-domain (OOD) in terms of both degradation types and language distribution.} The URGENT blind test set (URGENT 2025 \textit{test}) comprises real-world GSR data. The DNS 2020 test set (DNS \textit{test}) is a real-world dataset with noise and reverberation. Lastly, the AATC test set encompasses the widest range of degradation types, including secondary artifacts from imperfect neural network-based enhancements. 

\noindent\textbf{Baseline methods} We have selected various baseline methods to comprehensively show the effectiveness of our proposed method. We used two MetricGAN-based effective SE networks, MP-SENet~\cite{Lu_2023} and SEMamba~\cite{chao2025investigationincorporatingmambaspeech}. Although their initial task was denoising, results on recent challenges~\cite{zhang2025ccfaatc2025speech} also showed surprising performance in GSR settings. For LM-based approaches, we used LLaSE-G1~\cite{kang2025llaseg1incentivizinggeneralizationcapability} and MaskSR~\cite{li2024masksrmaskedlanguagemodel}. For non-LM generative methods, we used Universe++~\cite{scheibler2024universalscorebasedspeechenhancement}, SGMSE+~\cite{Richter_2023}, and ANYENHANCE~\cite{Zhang_2025}. Lastly, we employed VoiceFixer~\cite{Liu_2022} and USEMamba~\cite{chao2025universalspeechenhancementregression}. VoiceFixer~\cite{Liu_2022} is the first unified framework for high-fidelity speech restoration that uses a two-stage approach to restore degraded speech from multiple distortions. USEMamba~\cite{chao2025universalspeechenhancementregression} extended SEMamba to GSR via a hybrid flow matching~\cite{lipman2023flowmatchinggenerativemodeling} and regressive modeling approach. In our case, we utilized the version without flow matching, which achieved the best performance in non-blind phase results among other variants~\cite{chao2025universalspeechenhancementregression}. The loss is purely regressive towards minimizing sample point-wise distance without requiring GAN or any generative method.  
For all the baseline methods, we followed the official training codes. For LLaSE-G1, we used noise suppression mode, which may have degraded GSR performance since it only supports denoising and dereverberation. For USEMamba, because the code was unavailable, we implemented the model as described in the original paper.   

\noindent\textbf{Evaluation Metrics} 
We considered a range of neural and signal processing-based speech quality metrics. For neural speech quality prediction, we utilized SCOREQ~\cite{ragano2024scoreq}, UTMOS~\cite{saeki2022utmosutokyosarulabvoicemoschallenge}, and DNSMOS~\cite{reddy2021dnsmosnonintrusiveperceptualobjective} to assess the overall perceptual quality and naturalness. For SCOREQ, we used the version trained on synthetic data quality prediction in no-reference mode to better predict the quality of machine-generated speech. In DNSMOS, we utilized three distinct metrics: speech quality (SIG), background noise quality (BAK), and overall quality (OVRL) to provide a general evaluation. To evaluate signal fidelity, we used three metrics. Perceptual Evaluation of Speech Quality (PESQ)~\cite{941023} was included to calculate the signal similarity between the real and enhanced speech. Log spectral distance (LSD) was incorporated to quantify the spectral distortion of the restored speech against the clean speech. Lastly, Levenshtein phone similarity (LPS)~\cite{10363040} was used to capture speech mumbling and phoneme similarity, which is useful in evaluating generative speech enhancement models. We followed the official implementation~\cite{saijo2025interspeech2025urgentspeech} for calculating LPS. Across tables, the best and second-best scores are \textbf{boldfaced} and \underline{underlined}. Additionally, the real-time factor (RTF) was calculated on a single A6000 GPU to quantify the model's actual efficiency. RTF was calculated as the time required to process a second of speech.

\begin{table}[ht!]
\caption{Efficiency and objective performance evaluation results on AATC Challenge 2025. Official checkpoints have been utilized for models with $^*$.}
\label{tab:aatc-results}
\centering
\scriptsize
\renewcommand{\arraystretch}{0.94}
\setlength{\tabcolsep}{3pt}
\begin{tabular}{lcc|cccc}
\toprule
Model & \makecell{Params\\ (M)} & RTF & SIG & BAK & OVRL & PESQ \\
\midrule
Degraded & \multicolumn{2}{c|}{-}  & 2.89 & 3.02 & 2.47 & \textbf{1.91} \\
\midrule
MP-SENet & 2.3 & 0.022 &3.27 & 3.89 & 2.94 & 1.83 \\
SEMamba & 1.7 & 0.013 & 3.21 & 3.88 & 2.88 & 1.68 \\
USEMamba & 3.9 & 0.045 & \underline{3.36} & 3.99 & \underline{3.06} & 1.73 \\
Universe++  & 42.8 & 0.023 & 2.58 & 3.83 & 2.32 & 1.11 \\
Universe++$^*$ & 42.8 & 0.023 & 3.10 & 3.86 & 2.80 & 1.77 \\
LLaSE-G1$^*$ & 1072 & 0.011 & 3.15 & \underline{4.00} & 2.88 & 1.25 \\
\midrule
SEMamba++ (Ours) & 2.7 & 0.021 & \textbf{3.45} & \textbf{4.01} & \textbf{3.18} & \underline{1.84} \\
\bottomrule
\end{tabular}
\end{table}

\section{Experimental results}
\label{sec:expresults}

\subsection{GSR performance}
Table~\ref{tab:main-results} shows the experimental results for multiple backbone models. The table presents the objective evaluations for both the in-domain and OOD data. For in-domain data, we utilized VCTK-GSR \textit{test}, the test subset from VCTK simulated with multiple degradations. For OOD data, we utilized the validation and test subsets of the URGENT 2025 challenge. We set the sampling rate universally to 16 kHz. Most importantly, our model excels in most metrics across all datasets. Particularly on OOD datasets, our model exhibited the best performance by a substantial margin over the baselines. We provide further analysis on the performance improvement in Tables~\ref{tab:stagecount},~\ref{tab:unet}, and~\ref{tab:thoroughablation}. 

Overall, discriminative methods exhibit reasonable performance across both perceptual quality and signal fidelity metrics. In particular, SEMamba showed the best PESQ, LSD, and LPS in the VCTK-GSR \textit{test}. However, the LSD and LPS results in URGENT 2025 \textit{val} suggest that our proposed method shows competitive signal fidelity under data domain shift. Universe++, despite showing limited performance in PESQ, achieved competitive scores in perceptual quality metrics. 
Generative methods usually exhibit low LPS scores in both in-domain and OOD datasets, consistent with ~\cite{10363040} that generative speech enhancement exhibits unique errors such as phoneme substitution. Table~\ref{tab:aatc-results} shows the model complexity and objective evaluation results for the blind test set of AATC challenge 2025~\cite{zhang2025ccfaatc2025speech}. Considering that the dataset is OOD, comprising source data from different domains and distorted by unseen degradation types (Codec distortion and secondary artifacts), the results demonstrate generalization across various data domains and to unseen degradation types. Overall, our proposed method demonstrates noticeable restoration performance, regardless of those encountered during training. Tables~\ref{tab:main-results} and~\ref{tab:aatc-results} suggest that our model not only exhibits high fidelity on the seen dataset but also demonstrates significant generalization capacity on the unseen dataset or degradation types, with comparatively small parameter sizes. Thanks to the efficiency of frequency downsampling and the frequency GLP block, our model retains a lower RTF than most baselines. Surprisingly, despite its 1B parameters, LLaSE-G1 exhibited the lowest RTF, which is due to the sequence length compression utilizing codec techniques.

\begin{table}[t]
\caption{Joint denoising and reverberation performance, evaluated by DNSMOS on DNS 2020 test real recordings. All models are trained for solving general speech restoration. Diff. and Reg. denote the training with diffusion and regressive loss, respectively. $^*$ indicates models from official checkpoints, and $^\dagger$ represents model performance reported in the source papers.}
\label{tab:dns2020_result}
\centering
\scriptsize
\setlength{\tabcolsep}{1.5pt}
\renewcommand{\arraystretch}{0.94}
\begin{tabular}{lccc|ccc}
\toprule
Model & Methods & Params (M) & \makecell{Training data\\ (hours)}& SIG & BAK & OVRL \\
\midrule
Degraded  & \multicolumn{3}{c|}{-} & 3.053 & 2.509 & 2.255 \\
\midrule
ANYENHANCE$^{\dag}$   & MGM & 363.6 & 3.7k& \textbf{3.488} & 3.977 & 3.161 \\
MaskSR-M$^{\dag}$      & MLM & 145 & 800 & 3.430 & \textbf{4.025} & 3.136 \\
LLaSE-G1$^{\dag}$      & LM & 1072 & 500 & 3.472 & 3.996 & \underline{3.177}\\
VoiceFixer$^*$  & Reg. & 111.7 & 44 & 3.291 & 3.960 & 2.992 \\
USEMamba     & Reg. & 3.9 & 44& 3.239 & 3.937 & 2.923 \\ 
SGMSE+$^{\dag}$ & Diff. & 65.8 & 354 & 3.42 & 3.82 & 3.04 \\ 
Universe++  & Score & 42.8 & 44 & 2.646 & 3.756 & 2.374 \\
Universe++$^*$  & Score & 42.8 & 300& 3.077 & 3.681 & 2.708 \\
\midrule
SEMamba++ (Ours)& GAN & 2.7 & 44 & \underline{3.487} & \underline{4.020} & \textbf{3.206} \\ 

\bottomrule
\end{tabular}
\end{table}

\begin{figure*}[t]
    \centering
    \includegraphics[width=0.9\textwidth]{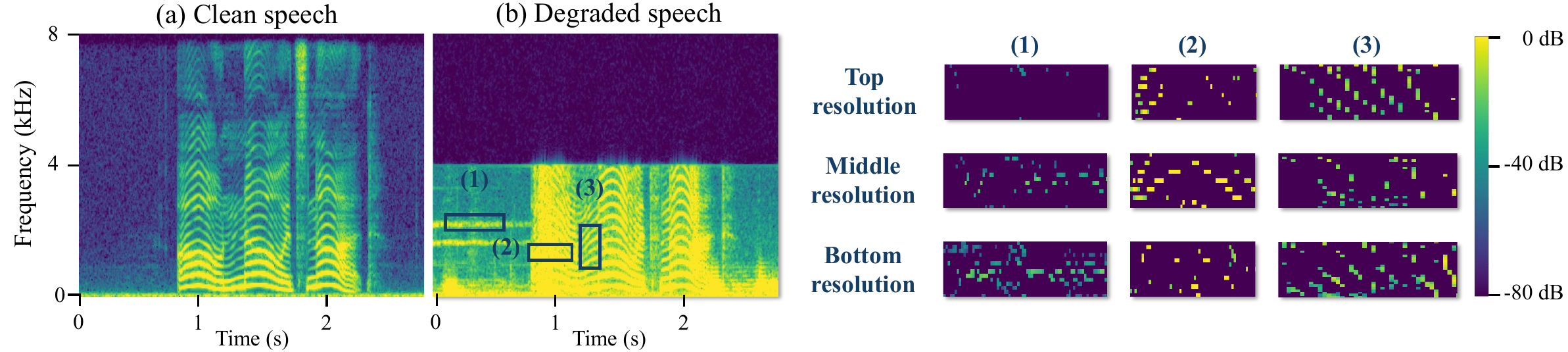}
    \caption{Gradient visualization of outputs from different branches in the proposed multi-branch method. (a) and (b) represent the magnitude spectrogram of the clean and the degraded speech, respectively. Resolution-wise visualizations of the gradient-weighted magnitude spectrogram are illustrated in (1), (2), and (3). The top resolution has a frequency dimension equal to $F'$.}
    \label{fig:gradbranch_vis}
\end{figure*}

\subsection{Task-wise performance}

We conducted a range of experiments to analyze how GSR models adapt to specific degradation types. Table~\ref{tab:dns2020_result} shows the joint denoising and dereverberation performance on the DNS 2020 test real recordings. For VoiceFixer, we utilized the official checkpoint. For ANYENHANCE, MaskSR-M, and LLaSE-G1, results are taken from \cite{Zhang_2025}, \cite{li2024masksrmaskedlanguagemodel}, and \cite{zhu2025meanflowseonestepgenerativespeech}, respectively. Among all GSR baselines, our proposed method achieves the highest overall quality score, outperforming both LLaSE-G1 and ANYENHANCE on the OVRL metric. Moreover, our method achieves the second-best performance in both SIG and BAK, indicating balanced performance. ANYENHANCE and MaskSR-M demonstrate strong results, thanks to the generalization capacity of generative methods. However, directly comparing these models with others may be inappropriate, since they are trained on a vastly larger dataset than the competing models. A thorough performance comparison using training data scaled to the same size is left for future work.  
VoiceFixer and USEMamba, trained on regressive objectives, deliver comparable performance in reducing background noise, as evidenced by the BAK score comparable to our proposed method and the masked modeling-based methods. However, the overall quality (OVRL) degrades due to weak speech quality (SIG). Although Universe++ shows good performance in the in-domain GSR setting of Table~\ref{tab:main-results}, its performance decreases in joint denoising and dereverberation. 

Furthermore, we provide Table~\ref{tab:efficiency_single_degrad_final} to conduct the intensity-wise analysis of degradations. Overall, SEMamba++ performed best across various intensities. In particular, for the most severe noise intensity (i.e., -15 dB), our model demonstrated robustness, indicating its real-world applicability across varying degrees of degradation. For bandwidth limitations, except at the 1 kHz cutoff frequency, our model achieved the best performance. 
MP-SENet and SEMamba, despite showing good denoising performance, cannot perform bandwidth extension because they rely on magnitude masking, which cannot generate arbitrary magnitude values in the high-frequency region. USEMamba, despite using magnitude mapping, performed poorly in the 1 kHz scenario, indicating that our proposed method is more flexible and effective. At 1 kHz, Universe++ achieves the highest UTMOS score, implying the strong potential of score-based generative modeling for unseen degradation intensity. 


\begin{table}[t]
\caption{UTMOS evaluation on VCTK-GSR \textit{test} dataset across different degradation intensities. The gray areas represent the evaluation results on unseen degradation intensities.} 
\label{tab:efficiency_single_degrad_final}
\centering
\scriptsize
\renewcommand{\arraystretch}{0.9}
\setlength{\tabcolsep}{3.5pt}
\begin{tabular}{l|cccc|cccc}
\toprule

\multirow{3}{*}{Model}
& \multicolumn{8}{c}{Degradation types}\\
\cmidrule(lr){2-9}
& \multicolumn{4}{c|}{Noise (dB)}
& \multicolumn{4}{c}{Bandwidth limitation (kHz)}\\
\cmidrule(lr){2-5}\cmidrule(lr){6-9}

& \cellcolor{lightgray}-15
& -10
& 0
& 10
& \cellcolor{lightgray}1
& 2
& 4
& 6\\

\midrule
Degraded
& \cellcolor{lightgray}1.42
& 1.46 & 1.6 & 2.25 
& \cellcolor{lightgray}1.96 & 2.75 & 3.73 & 3.95
\\

\midrule
MP-SENet
& \cellcolor{lightgray}1.94
 & 2.34 & 3.33 & 3.82
 & \cellcolor{lightgray}1.85 & 2.80 & 3.73 & 3.99
\\

SEMamba
& \cellcolor{lightgray}2.06 
& 2.51 & \underline{3.48} & \underline{3.89}
& \cellcolor{lightgray}1.85 & 2.79 & 3.72 & 3.97
\\ 



USEMamba
& \cellcolor{lightgray}1.98 
& 2.38 & 3.35 & 3.83 
& \cellcolor{lightgray}1.52
& 2.89 & \underline{3.85} & \underline{4.04}
\\

Universe++ 
& \cellcolor{lightgray}\underline{2.38} 
& \underline{2.85}& 3.41&3.59 
& \cellcolor{lightgray}\textbf{3.33} 
& \underline{3.57}& 3.66& 3.68
\\


LLaSE-G1$^*$
& \cellcolor{lightgray}2.26 
& 2.41 & 3.24 & 3.46 
& \cellcolor{lightgray}1.69 & 2.14 & 2.86 & 3.3
\\

\midrule
SEMamba++ (Ours)
& \cellcolor{lightgray}\textbf{2.90}
& \textbf{3.34} & \textbf{3.85} & \textbf{4.00}
& \cellcolor{lightgray}\underline{2.94} & \textbf{3.86} & \textbf{4.07} & \textbf{4.09} \\ 

\bottomrule
\end{tabular}
\end{table}
\vspace{-0.5em}


\subsection{Multi-resolution gradient analysis}
\label{sec:mranalysis}
To demonstrate that the proposed multi-resolution parallel TFDP processing encourages different resolutions to capture various patterns of the input spectrogram, we visualized the input attribution maps for the outputs of each resolution in the first layer (Figure~\ref{fig:gradbranch_vis}). Following established gradient-based visualization techniques~\cite{Selvaraju_2019}, we computed the gradients of the outputs of different resolutions with respect to the input magnitude spectrogram and then applied a ReLU to the gradients to focus on the TF bins that contributed positively to each resolution's output. The gradients are then normalized to [0, 1], and only the largest 10\% values are retained, with the remaining values masked to 0, producing sparse attribution masks. We refer to these gradients as \textbf{influential gradients}. Lastly, we applied element-wise multiplication of the gradients to the degraded spectrogram to derive the gradient-weighted magnitude spectrogram. This highlights the TF regions most influential in deriving outputs in each resolution. As shown in (a) and (b), the input spectrogram exhibits distortions from noise, reverberation, lowpass filtering, and clipping. As illustrated in (1), bottom resolution exhibits the strongest response to noise patterns, while top resolution shows relatively weaker noise sensitivity. The middle resolution effectively captures the speech patterns in (2). Lastly, in (3), the top resolution shows the largest response to the harmonic patterns. Note that (3) is rotated $90^\circ$ counterclockwise. Collectively, these observations demonstrate that each resolution specializes in distinct spectro-temporal patterns, operating as complementary modules that together provide comprehensive feature extraction. 

To further analyze the diverse feature extraction of the proposed MR-parallel TFDP, influential gradients were computed at each resolution, and the Intersection-over-Union (IoU) was measured to quantify the distribution of gradients across different resolutions. A lower IoU indicates that the spectrogram regions with influential gradients differ more significantly between resolutions, suggesting more diverse spectral modeling. In Table~\ref{tab:parallelvssequential}, the IoU scores obtained from MR-parallel and MR-sequential were compared using a Wilcoxon signed-rank test~\cite{wilcoxon1945individual} with 3.9k samples to assess statistical differences. The parallel TFDP processing yields significantly lower IoU values than its sequential counterpart ($p<0.05$), indicating that the parallel design encourages more diverse and complementary spectral modeling across multiple resolutions.

\begin{table}[t]
\centering
\scriptsize
\caption{Intersection Over Union (IoU) values of influential gradients across multiple resolutions.}
\label{tab:parallelvssequential}
\begin{tabular}{lccc}
\toprule
Method & VCTK-GSR \textit{test} & URGENT \textit{test} & DNS \textit{test} \\
\midrule
MR-Parallel & 0.036 & 0.036 & 0.037 \\
MR-Sequential & 0.045 & 0.047 & 0.046  \\
\bottomrule
\end{tabular}
\end{table}


\begin{figure}[htbp]
    \centering
    \includegraphics[width=0.45\textwidth]{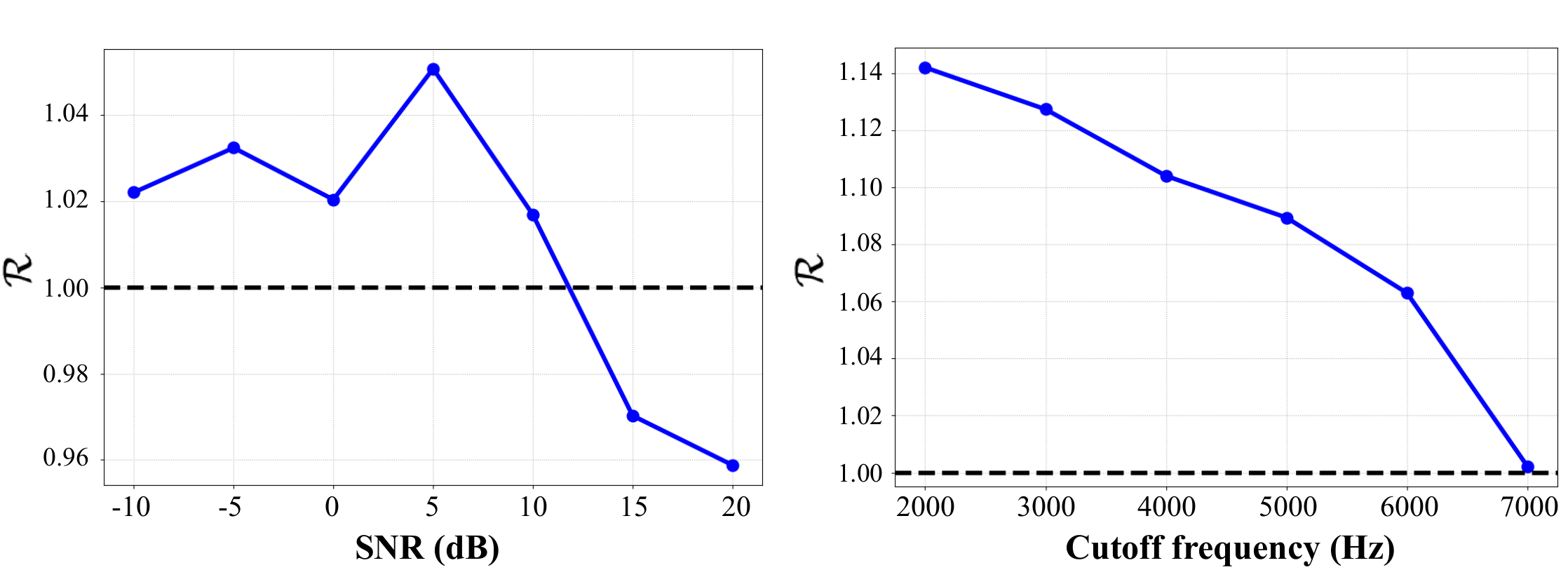}
    \caption{The ratio of gradient norms under different degradation types and intensity. $\mathcal{R} > 1$ indicates the larger contribution of the Global Periodicity module than the Local module.}
    \label{fig:glocal_ratio}
\end{figure}

\vspace{-0.5em}

\subsection{Intensity-wise analysis on GLP module contribution}
To analyze how the frequency GLP block adjusts to different degradations, we computed the gradient norms of the GP and L modules. For degradations, we chose additive noise with SNR ranging from -10 to 20 dB and bandwidth limitation with a cutoff frequency ranging from 2000 to 7000 Hz. The entire training dataset was used for the analysis. First, for input speech with varying degradation levels, we calculated the gradients of the predicted speech with respect to GP and L modules' outputs in the top resolution, each with dimensions $T\times F'\times C$. For each degradation level, we computed the mean $ L_2$ Norm of the gradients across all layers to obtain the gradient norms $G_{\text{GP}}$ and $G_{\text{L}}$. The ratio of gradient norms was then defined as $\mathcal{R}=G_{\text{GP}}/{G_{\text{L}}}$. $\mathcal{R}$ greater than 1 means that the contribution of the GP module is larger than that of the L module for a certain degradation. Figure~\ref{fig:glocal_ratio} illustrates the result, which explicitly quantifies the relative contribution of the GP module compared to the L module under different degradation conditions. For noise, high intensity (low SNR) results in $\mathcal{R} > 1$, implying dominance of the GP module, whereas low noise intensity (high SNR) favors the L module. This contrast is intuitive because noise patterns generally span the entire frequency axis, whereas speech patterns do not. Thus, the GP module effectively extracts the relevant features when the noise intensity is high, whereas the reverse is true when the intensity is low. For bandwidth extension, the ratio always favors the GP module, and the gap decreases as the cutoff frequency increases. When the lowpass filter is applied, the L module is ineffective at high frequencies because neighboring bins contain no meaningful information. As the cutoff frequency increases, the ratio approaches 1 because the size of the missing high-frequency region decreases. Further experimental results are provided in Table~\ref{tab:stagecount} to justify the effectiveness of the GP module.

\begin{figure}[t]
    \centering
    \includegraphics[width=0.41\textwidth]{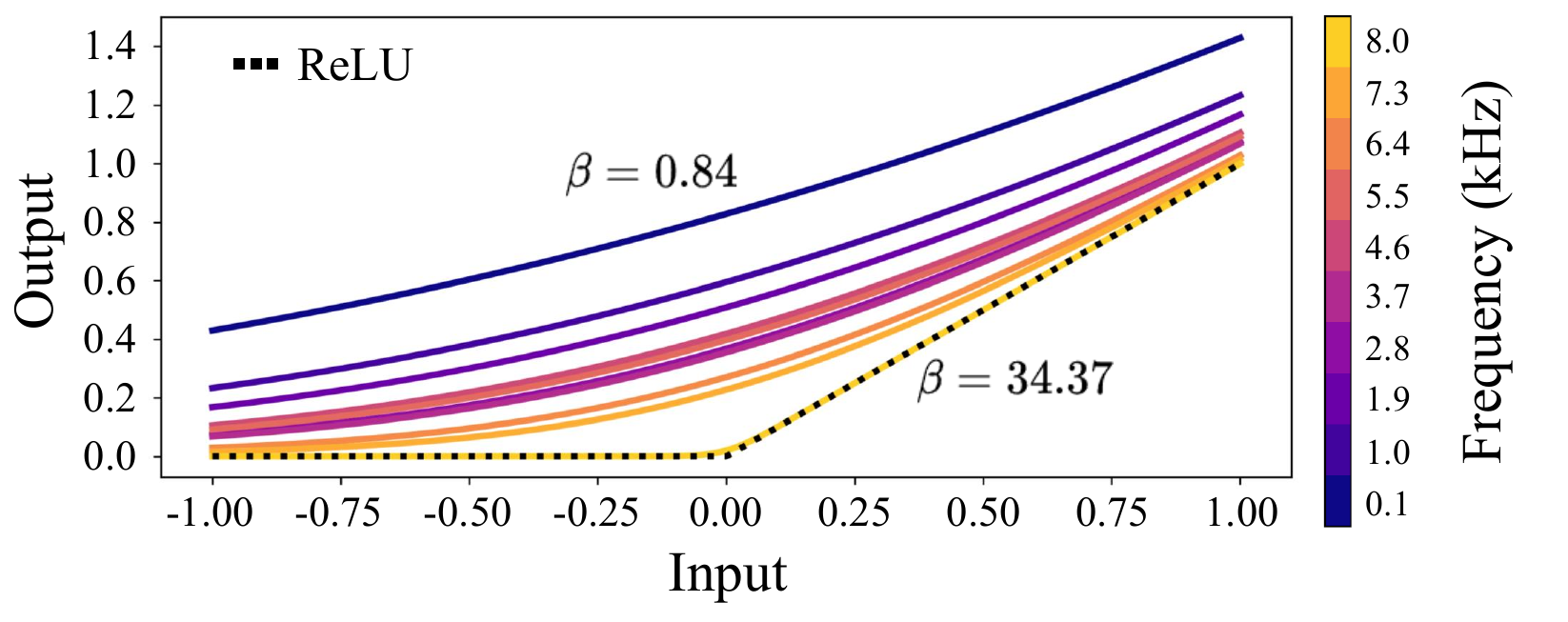}
    \caption{Softplus functions with the learnable $\beta$ for each frequency bin. The black dotted line denotes the ReLU function.}
    \label{fig:learnsoftplus}
\end{figure}
\subsection{Learnable softplus visualization}

In Figure~\ref{fig:learnsoftplus}, we have visualized the softplus function with learnable $\beta$ for each frequency bin used for the magnitude mapping. We selected 10 frequency values between 0.1 kHz and 8 kHz and visualized the corresponding softplus functions using their $\beta$ values. Generally, $\beta$ in low-frequency bands is smaller than that in high-frequency bands. The resulting softplus functions hold larger responses in low-frequency bands. At the Nyquist frequency (8 kHz), the value of $\beta$ is 34.4, thereby shaping softplus to behave like a ReLU. This trend implies that the overall energy in the low-frequency region is higher than in the high-frequency region in clean speech. Overall, the figure suggests that frequency-aware $\beta$ adjustments are effective, showing the effectiveness of the proposed learnable softplus.


\subsection{Ablation studies}
We have designed ablations for the proposed Frequency GLP, multi-resolution parallel TFDP processing block, learnable mapping, and vocoder-style training. All ablations include evaluation on both the seen (VCTK-GSR \textit{test}) and unseen (URGENT 2025-\textit{test} and DNS-\textit{test}) datasets.

\subsubsection{On Frequency GLP}
\begin{table}[t]
\caption{Comparison and ablation studies on various frequency mixing modules using SEMamba++ as a backbone. }
\label{tab:stagecount}
\centering
\scriptsize
\renewcommand{\arraystretch}{0.9}
\setlength{\tabcolsep}{1.0pt}
\begin{tabular}{lcc|cccccc}
\toprule
\multirow{3}{*}{Design}
& \multirow{3}{*}{\makecell{Params \\ (M)}} & \multirow{3}{*}{RTF} & \multicolumn{2}{c|}{In-domain}
& \multicolumn{4}{c}{Out-of-domain} \\
& &  & \multicolumn{2}{c}{VCTK-GSR \textit{test}}
& \multicolumn{2}{c}{URGENT \textit{test}}
& \multicolumn{2}{c}{DNS \textit{test}} \\

\cmidrule(lr){4-5}
\cmidrule(lr){6-7}
\cmidrule(lr){8-9}

& &  & UTMOS & OVRL 
& UTMOS  & OVRL 
& UTMOS & OVRL \\
\midrule
Mamba & 2.7& 0.025 & \textbf{3.55} & 3.12 & 2.50 & 3.04 & 2.70 & 3.06 \\ 
Transformer & 2.9 & 0.027 & 2.95 & 2.95 & 2.24 & 2.95 & 2.40 & 3.04 \\
Conformer & 2.4 & 0.028 & 3.45 & 3.06 & 2.46 & 3.04 & 2.93 & 3.17 \\
SpatialNet & 2.6 & 0.024 & 3.46 & 3.10 & 2.36 & 2.98 & 2.68 & 3.11 \\
\midrule
GLP (Ours)& 2.7 & 0.021 & \textbf{3.55} & \textbf{3.14} & \textbf{2.61} & \textbf{3.13} & \textbf{3.02} & \textbf{3.21} \\  
\midrule
\textit{w/o} GP module & 2.9 & 0.030 & 3.43 & 3.08  & 2.51 & 3.05 & 2.64 & 3.09\\
Serial& 2.6 & 0.020 & 3.52 & 3.1 & 2.45 & 2.99 & 2.82 & 3.11 \\
FAN$\shortto$Linear& 2.8 & 0.019 & 3.45 & 3.10 & 2.50 & 3.05 & 2.96 & 3.17 \\

\bottomrule
\end{tabular}
\end{table}

In Table~\ref{tab:stagecount}, we present an analysis showing the effectiveness of our proposed frequency GLP. By default, all experiments examined in this study shared the same settings, including time Mamba, multi-resolution parallel architecture, vocoder-style training, and learnable mapping. We first present four common choices for frequency feature extraction modules: Mamba~\cite{gu2024mambalineartimesequencemodeling}, Transformer, Conformer, and SpatialNet. We have made minor modifications to align parameter sizes across modules. For Mamba, we increased the expansion factor from 2 to 4. For SpatialNet, we have repeated the cross-band block twice. We have also conducted ablations of different design choices in our proposed method. We first replaced the GP module with the L module, thereby removing the global and periodic frequency interaction capacity; we denote this as \textbf{\textit{w/o} GP module}. We also tested the serial connection of L and GP modules (local$\rightarrow$global) to verify whether the selectivity in our proposed method not only makes the model explainable but also, most importantly, enhances performance. We denote this as \textbf{Serial}. Lastly, we replaced FAN with a simple linear layer to assess whether capturing periodicity was essential, denoted as \textbf{FAN$\rightarrow$Linear}. As shown in Table~\ref{tab:stagecount}, the proposed GLP block achieves the best performance among the baselines on both the seen and unseen datasets. The performance gap is most evident on the unseen datasets, indicating GLP's strong generalization ability relative to other advanced frequency feature extraction solutions (Mamba, Transformer, Conformer, and SpatialNet). Mamba models global features but is sensitive to the sequential order of features, which is less aligned with the characteristics of the frequency features, where the absolute spectral position holds important information (e.g., f0). 
Transformer~\cite{lee_2024}, given the bias-free nature of attention, assumes no inductive biases for the input features. This nature, while enhancing generalization across different kinds of input, could be advanced if necessary biases are incorporated. Moreover, the RTF results indicate that our proposed method is the most efficient among the baselines, owing to the high efficiency of the linear operation. The ablation results show that all components of GLP contributed to improving performance. The most critical component in GLP blocks is the GP module (see \textit{w/o} GP module), suggesting the importance of capturing both global and periodic structures in frequency bins through linear combination. The result also indicates that using the GP module yields significant gains in inference efficiency. Selective information fusion outperformed serial connections, with the largest gap observed on the unseen datasets. Lastly, substituting FAN with a simple linear layer also led to performance degradation, underscoring the importance of capturing periodicity. It is worth noting that the performance gap on the DNS \textit{test} was the smallest, suggesting that the importance of capturing periodicity is greater in bandwidth extension than in denoising or dereverberation.

\subsubsection{On multi-resolution parallel TFDP processing}
\begin{table}[t]
\caption{Ablation studies on the proposed multi-resolution parallel TFDP processing. $\times$2 and $\times$4 denote downsampling by factors of 2 and 4, respectively.}
\label{tab:unet}
\centering
\scriptsize
\setlength{\tabcolsep}{0.7pt}
\begin{tabular}{lcc|cccccc}
\toprule
\multirow{3}{*}{Design}
& \multirow{3}{*}{\makecell{Params \\ (M)}} & \multirow{3}{*}{RTF} & \multicolumn{2}{c|}{In-domain}
& \multicolumn{4}{c}{Out-of-domain} \\
& & & \multicolumn{2}{c}{VCTK-GSR\,\textit{test}}
& \multicolumn{2}{c}{URGENT\,\textit{test}}
& \multicolumn{2}{c}{DNS\,\textit{test}} \\

\cmidrule(lr){4-5}
\cmidrule(lr){6-7}
\cmidrule(lr){8-9}

&  & & UTMOS & OVRL 
& UTMOS  & OVRL 
& UTMOS & OVRL \\
\midrule  
\makecell[l]{MR-Parallel\\(Ours)} & 2.7 &  0.021 & \textbf{3.55} & \textbf{3.14} & \textbf{2.61} & \textbf{3.13} & 3.02 & \textbf{3.21} \\  
\midrule
MR-Sequential & 2.7 & 0.021 & 3.54 & 3.12  & 2.53 & 3.08 & 2.91 & \textbf{3.21}\\
SR & 9.0 & 0.034 & 3.53 & 3.11  & 2.42 & 2.85 & 2.19 & 2.48\\
\textit{w/o} down. & 2.8 & 0.028 & 3.50 & 3.09 & 2.44 & 2.95 & 2.72 & 3.0 \\
T$\times$4 & 2.6 & 0.019 & 3.39 & 3.04 & 2.46 & 3.03 & 2.75 & 3.15 \\
T$\times$2 F$\times$2& 2.7 & 0.019 & \textbf{3.55} & 3.10 & 2.53 & 3.03 & \textbf{3.03} & 3.14 \\

\bottomrule
\end{tabular}
\end{table}

In Table~\ref{tab:unet}, we analyzed the effectiveness of the proposed multi-resolution parallel TFDP block. First, we examined the multi-resolution (MR) sequential TFDP processing at each resolution to assess the effectiveness of the parallel TFDP processing, denoted as \textbf{MR-Sequential}. We also designed a variant that uses a single resolution (denoted \textbf{SR}). The channel dimension $C$ in the single-resolution was set to 144 to na\"ively widen the model to increase its expressivity. Regarding frequency-only downsampling, we provided three variants: one without downsampling, one with time-only downsampling, and one with downsampling on both axes. We denote these experiments as \textbf{\textit{w/o} down.}, \textbf{T$\times$4}, and \textbf{T$\times$2 F$\times$2}, respectively. We observe from Table~\ref{tab:unet} that the key components proposed in our method improve performance and generalization. A modification from the sequential to the parallel resulted in an overall performance gain. Together with the analysis results in Section~\ref{sec:mranalysis}, we assume that the diverse spectral modeling capacity of parallel processing brought performance improvements. The single-resolution result showed a slight decrease in the in-domain dataset but a huge drop in the OOD dataset, suggesting the possibility of overfitting. Time-only downsampling showed worse performance than \textbf{T$\times$2 F$\times$2}, suggesting that the upsampling artifacts incurred by frequency downsampling are more ignorable than those from time downsampling. Surprisingly, a variant without downsampling (\textit{w/o} ds) showed a slight drop in every metric, implying that frequency-only downsampling is not only efficient but also effective in helping the model capture patterns of multiple resolutions.

\subsubsection{From SEMamba to SEMamba\texttt{++}}
\begin{table}[t]
\caption{Ablation studies on the model design choices. LMask., LMap., and Metric. refer to learnable masking, mapping, and MetricGAN, respectively.}
\label{tab:thoroughablation}
\centering
\scriptsize
\setlength{\tabcolsep}{0.2pt}
\begin{tabular}{lcc|cccccc}
\toprule
\multirow{3}{*}{Design}
& \multirow{3}{*}{\makecell{Params \\ (M)}} & \multirow{3}{*}{RTF} & \multicolumn{2}{c|}{In-domain}
& \multicolumn{4}{c}{Out-of-domain} \\
& &  & \multicolumn{2}{c}{VCTK-GSR \textit{test}}
& \multicolumn{2}{c}{URGENT \textit{test}}
& \multicolumn{2}{c}{DNS \textit{test}} \\

\cmidrule(lr){4-5}
\cmidrule(lr){6-7}
\cmidrule(lr){8-9}

& &  & UTMOS & PESQ
& UTMOS  & OVRL 
& UTMOS & OVRL \\
\midrule
SEMamba & 1.7 & 0.013 & 2.34 & 2.13 & 1.79 & 2.80 & 1.57 & 2.75 \\
LMask.$\shortto$ LMap. & 1.7 & 0.013 & 2.30 & \textbf{2.22} & 1.82 & 2.87 & 1.82 & 2.90 \\
\hspace{0.1em} Metric.$\shortto$Vocoder & 1.7 & 0.013 & 3.32 & 1.71 & 2.28 & 2.91 & 2.14 & 2.88 \\
\hspace{0.6em} TFMamba-large & 2.7 & 0.018 & 3.31 & 1.75 & 2.23 & 2.84 & 2.31 & 2.85 \\
\midrule
\hspace{0.6em} Ours & 2.7 & 0.021 & \textbf{3.55} & 1.77 & \textbf{2.61} & \textbf{3.13} & \textbf{3.02} & \textbf{3.21}\\

\bottomrule
\end{tabular}
\end{table}

Table~\ref{tab:thoroughablation} denotes the component-wise analysis of SEMamba++ with regard to SEMamba. The transition from learnable masking (LMask.) to learnable mapping (LMap.) yielded overall performance gains, particularly on unseen data. Vocoder-style training with LSGAN, MRD, and MS-SB-CQTD improved perceptual quality on all datasets, though it resulted in a significant drop in PESQ, due to MetricGAN objective explicitly optimizing PESQ. Lastly, modifying the bottleneck design from the original TFMamba in SEMamba to our proposed method consistently improved all evaluation metrics across both in-domain and OOD scenarios, with the largest gains in the latter. TFMamba, with the parameter size (TFMamba-large) equivalent to our proposed method by adjusting $C$ from 64 to 80, has also been included in the experiment. These results suggest that, for the same training loss functions, our proposed method is architecturally superior to TFMamba and TFMamba-large.

\section{Conclusions}
\label{sec:conclusions}

This work proposed SEMamba++, an effective and efficient solution for general speech restoration that improves upon SEMamba by utilizing speech-specific features as inductive biases. The proposed model comprises two novel building blocks: the Frequency GLP and the multi-resolution parallel TFDP block. The Frequency GLP captures global, local, and periodic patterns in spectral data, yielding superior performance and greater efficiency than modern frequency modeling modules. The multi-resolution parallel TFDP block introduces resolution-wise complementary feature extraction, enabling the module to capture diverse features with greater expressivity. Moreover, we introduce a learnable softplus-based mapping that effectively adjusts frequency response to better model the speech spectrum. Our model achieves the best overall quality across one in-domain dataset and four OOD datasets, with only 2.7M parameters. 

\section{Limitations}

The proposed Frequency GLP may not be directly deployed in scenarios requiring sampling-frequency-independent processing due to the linear operation applied directly to the frequency axis. More advanced training objectives that improve both perceptual quality and signal fidelity metrics in the in-domain dataset have been underexplored in this work. 

{\footnotesize

\section{Use of Generative AI Disclosure}

Generative AI has been used to revise manuscripts, including correcting grammar.  

\section{Acknowledgements}

This work was supported by the National Research Foundation of Korea (NRF) grant (No. RS-2024-00337945), the STEAM research grant (No. RS-2024-00464269) funded by the Ministry of Science and ICT of Korea government (MSIT), the BK21 FOUR program through the NRF grant funded by the Ministry of Education of Korea government (MOE), and Industrial Technology Innovation R\&D program of MOTIE/KEIT. [RS-2025-24533624, Development of SoC and Ultrasound-Based AI High-Speed Gas Leak Detection Technology for Preventing Gas Explosions and Human Casualties].

\bibliographystyle{IEEEtran}
\bibliography{mybib}
}

\end{document}